\documentclass{article}
\usepackage{spconf,amsmath,graphicx}

\usepackage{enumitem}
\setlist{nosep, leftmargin=14pt}

\usepackage{mwe} 

\title{ViG-UNet: Vision Graph Neural Networks for Medical Image Segmentation}
\name{Juntao Jiang\textsuperscript{1}, Xiyu Chen\textsuperscript{2}, Guanzhong Tian\textsuperscript{3}* and Yong Liu \textsuperscript{1}*\thanks{*Corresponding authors}}
\address{1. College of Control Science and Engineering, Zhejiang University, Hangzhou, China \\ 2. Polytechnic Institute, Zhejiang University, Hangzhou, China\\3. Ningbo Innovation Center, Zhejiang University, Ningbo, China}
\begin{document}
%
\maketitle

\begin{abstract}
Deep neural networks have been widely used in medical image analysis and medical image segmentation is one of the most important tasks. U-shaped neural networks with encoder-decoder are prevailing and have succeeded greatly in various segmentation tasks. While CNNs treat an image as a grid of pixels in Euclidean space and Transformers recognize an image as a sequence of patches, graph-based representation is more generalized and can construct connections for each part of an image. In this paper, we propose a novel ViG-UNet, a graph neural network-based U-shaped architecture with the encoder, the decoder, the bottleneck, and skip connections. The downsampling and upsampling modules are also carefully designed. The experimental results on ISIC 2016, ISIC 2017 and Kvasir-SEG datasets demonstrate that our proposed architecture outperforms most existing classic and state-of-the-art U-shaped networks.
\end{abstract}


\section{Introduction}
Recent years have witnessed the rise of deep learning and its broader applications in computer vision tasks. As one of the most heated topics of computer vision applied in medical scenarios, image segmentation, identifying the pixels of organs or lesions from the background, plays a crucial role in computer-aided diagnosis and treatment, improving efficiency and accuracy.

Currently, medical image segmentation methods based on deep learning mainly use fully convolutional neural networks (FCN) with U-shaped encoder-decoder architecture such as U-Net \cite{ronneberger2015u} and its variants. Composed of a symmetric encoder-decoder with skip connections, U-Net uses convolutional layers and downsampling modules for feature extraction, while convolutional layers and upsampling modules for pixel-level semantic classification. The skip connection operation can maintain spatial information from a high-resolution feature, which may be lost in downsampling. Following this work and based on a fully convolutional structure, a lot of U-Net's variants like Attention-UNet \cite{oktay2018attention}, UNet++  \cite{zhou2018unet++} and so on, have been proposed and achieved great success. Recently, as Transformer-based methods like ViT \cite{dosovitskiy2020image} achieved good results in image recognition tasks, thanks to their capability of enhancing global understanding of images, extracting information from the inputs and their interrelations, the Transformer-based medical image segmentation models such as Trans-UNet \cite{chen2021transunet} and Swin-UNet \cite{cao2021swin} also have been proposed and showed competitive performance.

While CNNs treat an image as a grid of pixels in Euclidean space and Transformer recognizes an image as a sequence of patches, graph-based representation can be more generalized and reflect the relationship of each part in an image. Since the graph neural network (GNN) \cite{gori2005new} was first proposed, the techniques for processing graphs have been researched a lot. A series of spatial-based GCNs \cite{micheli2009neural, niepert2016learning} and spectral-based GCNs \cite{bruna2013spectral, henaff2015deep, defferrard2016convolutional, kipf2016semi} are widely proposed and applied. In recent work, Han et al. \cite{han2022vision} proposed a Vision GNN (ViG), which splits the image into many blocks regarded as nodes and constructs a graph representation by connecting the nearest neighbors, then uses GNNs to process it. It contains Grapher modules with graph convolution to aggregate and update graph information and  Feed-forward Networks (FFNs) modules with fully connected networks for node feature transformation, which performed well in image recognition tasks. ViG-S has achieved 0.804 Top-1 accuracy and ViG-B has achieved 0.823 on ImageNet \cite{krizhevsky2017imagenet}.

Motivated by the success of ViG model, we propose a ViG-UNet to utilize the powerful functions of ViG for 2D medical image segmentation in this work. The graph-based representation can also be effective in segmentation tasks. ViG-UNet is a GNN-based U-shaped architecture consisting of the encoder, the bottleneck and the decoder, with skip connections. We do comparison experiments on ISIC 2016 \cite{gutman2016skin}, ISIC 2017 \cite{codella2018skin} and Kvair-SEG \cite{jha2020kvasir} datasets.The results show that the proposed model outperformed most existing classic and state-of-the-art methods. The code will be released at \textit{https://github.com/juntaoJianggavin/ViG-UNet}.
\section{Methods}

\subsection{Architecture Overview}
ViG-UNet is a U-shape model with symmetrical architectures, whose architecture can be seen in Figure 1. It consists of structures of the encoder, the bottleneck, the decoder and skip-connections. The basic modules of ViG-UNet are the stem block, Grapher Modules, Feed-forward Networks (FFNs), downsampling and upsampling modules. The detailed settings of ViG-UNet can be seen in Table 1, where $D$ means feature dimension, $E$ means the numbers of convolutional layers in FFNs, $K$ means the number of neighbors in GCNs, $H \times W$ means the output image size.
\begin{figure*}[htbp]
\centering
\centerline{\includegraphics[width=14.01cm]{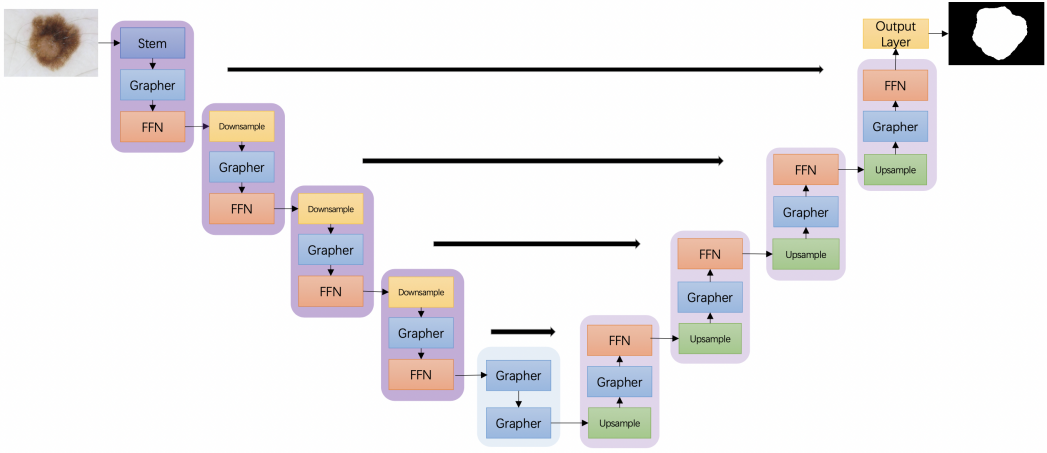}}
\caption{The architecture of ViG-UNet: the basic modules are the stem block for visual embedding, Grapher Modules, Feed-forward Networks (FFNs) modules, downsampling modules in the encoder and upsampling modules in the decoder.}
\end{figure*}

\begin{table}
\caption{Detailed settings of ViG-UNet}
\label{comp}
\center
\renewcommand{\arraystretch}{1.3}

\begin{tabular}{c|c|c}
\hline
Module                         &   Output size     & Architecture    \\ \hline
Stem & 
$\frac{H}{2} \times \frac{W}{2}$ & Conv×2 \\
Grapher + FFN &$\frac{H}{2} \times \frac{W}{2}$ & $\left[\begin{array}{c}D=32 \\ E=2 \\ K=9\end{array}\right]$ \\

Downsampling &$\frac{H}{4} \times \frac{W}{4}$ & Conv \\

Grapher + FFN &$\frac{H}{4} \times \frac{W}{4}$ & $\left[\begin{array}{c}D=64\\ E=2 \\ K=9\end{array}\right]$ \\

Downsampling &$\frac{H}{8} \times \frac{W}{8}$ & Conv \\

Grapher + FFN &$\frac{H}{8} \times \frac{W}{8}$ & $\left[\begin{array}{c}D=128\\ E=2 \\ K=9\end{array}\right]$ \\

Downsampling &$\frac{H}{16} \times \frac{W}{16}$ & Conv \\

Grapher + FFN &$\frac{H}{16} \times \frac{W}{16}$ & $\left[\begin{array}{c}D=256\\ E=2 \\ K=9\end{array}\right]$ \\

Downsampling &$\frac{H}{32} \times \frac{W}{32}$ & Conv \\

Grapher $\times$2 &$\frac{H}{32} \times \frac{W}{32}$ & $\left[\begin{array}{c}D=512 \\ K=9\end{array}\right] \times 2$ \\

Upsampling &$\frac{H}{16} \times \frac{W}{16}$ & bilinear + Conv \\

Grapher + FFN &$\frac{H}{16} \times \frac{W}{16}$ & $\left[\begin{array}{c}D=256\\ E=2 \\ K=9\end{array}\right]$ \\

Upsampling &$\frac{H}{8} \times \frac{W}{8}$ & bilinear + Conv \\

Grapher + FFN &$\frac{H}{8} \times \frac{W}{8}$ & $\left[\begin{array}{c}D=128\\ E=2 \\ K=9\end{array}\right]$ \\

Upsampling &$\frac{H}{4} \times \frac{W}{4}$ & bilinear + Conv \\

Grapher + FFN &$\frac{H}{4} \times \frac{W}{4}$ & $\left[\begin{array}{c}D=64\\ E=2 \\ K=9\end{array}\right]$ \\

Upsampling &$\frac{H}{2} \times \frac{W}{2}$ & bilinear + Conv \\

Grapher + FFN &$\frac{H}{2} \times \frac{W}{2}$ & $\left[\begin{array}{c}D=32\\ E=2 \\ K=9\end{array}\right]$ \\

Final Layer &$H \times W$ & bilinear + Conv \\\hline

\end{tabular}
\end{table}

\subsection{Stem Block, Upsampling, Downsampling and Skip-Connections}
In the stem block, two convolutional layers are used with stride 1 and stride 2, respectively. The output features have height and width equal to $\frac{H}{2}$ and $\frac{W}{2}$ , where $H$, $W$ are the original height and width of the input image. And the position embedding is added.
We used a convolutional layer with stride 2 for downsampling operation and a bilinear for upsampling operation with the scale factor 2 following with a convolutional layer. The output of each FFN in the encoder is added to the output of the FFN in the decoder.

\subsection{Grapher Module}
Vision GNN first builds an image's graph structure by dividing it into $N$ patches, converting them into feature vectors, and then recognizing them as a set of nodes $\mathcal{V}=\left\{v_1, v_2, \cdots, v_N\right\}$. A $K$ nearest neighbors method is used to find $K$ nearest neighbors $\mathcal{N}\left(v_i\right)$ for each node $v_i$. An edge $e_{j i}$ is added from $v_j$ to $v_i$ for all $v_j \in \mathcal{N}\left(v_i\right)$. In this way, a graph representation of an image $\mathcal{G}=(\mathcal{V}, \mathcal{E})$ is obtained, where $e_{j i} \in  \mathcal{E}$. For the constructed graph representation $\mathcal{G}=G(X)$ and the input feature ${x}_i$, the aggregation operation calculates the representation of a node by aggregating features of neighboring nodes. Then the update operation merge the aggregated feature. The updated feature $\mathbf{x}_i^{\prime}$ can be represented as:
\begin{equation}
\mathbf{x}_i^{\prime}=h\left(\mathbf{x}_i, g\left(\mathbf{x}_i, \mathcal{N}\left(\mathbf{x}_i\right); W_{aggregate}\right);W_{\text {update }}\right),
\end{equation}
where $W_{aggregate}$ and $W_{update}$ are the learnable weights of the aggregation and update operations. 
\begin{equation}
g(\cdot)=\mathbf{x}_i^{\prime \prime}=\left[\mathbf{x}_i, \max \left(\left\{\mathbf{x}_j-\mathbf{x}_i \mid j \in \mathcal{N}\left(\mathbf{x}_i\right)\right\}\right]\right.,
\end{equation} and
\begin{equation}
h(\cdot)=\mathbf{x}_i^{\prime}=\mathbf{x}_i^{\prime \prime} W_{\text {update }}+b_h,
\end{equation}
$b_h$ is the bias. And following the design of original ViG networks, the $g(\cdot)$ operation uses the max-relative graph convolution \cite{li2019deepgcns}.

The aggregated feature $\mathbf{x}_i^{\prime \prime}$ is split into $h$ heads, then each head is updated with different weights. Then the updated feature $\mathbf{x}_i^{\prime}$ can be obtained by concatenating all the heads:

\begin{equation}
\begin{aligned}
&\mathbf{x}_i^{\prime}=[\mathbf{x}_{i\text {head1 }}^{\prime \prime} W_{\text {update }}^1+b_{h 1}, \mathbf{x}_{i\text {head } 2}^{\prime \prime} W_{\text {update }}^2+b_{h 2}, \\
&\mathbf{x}_{i\text {head } 3}^{\prime \prime} W_{\text {update }}^3+b_{h 3},\cdots\mathbf{x}_{i\text {headh }}^{\prime \prime} W_{\text {update }}^h+b_{h h}]
\end{aligned}
\end{equation}
where $\mathbf{x}_{i\text {head1 }}^{\prime \prime},\mathbf{x}_{i\text {head2 }}^{\prime \prime},\cdots,\mathbf{x}_{i\text {headh }}^{\prime \prime}$ represent the split heads from  $\mathbf{x}_i^{\prime}$, $W_{\text {update }}^1,W_{\text {update }}^2,\cdots,W_{\text {update }}^h$ represent different weights and $b_{h 1},b_{h 2},\cdots,b_{h h}$ represent different biases.

For the input feature $X$, the output feature $Y$ after a Grapher module can be represented as:
\begin{equation}X_1=XW_{in}+b_{in},\end{equation}
\begin{equation}Y=\operatorname{Droppath}(\operatorname{GELU}(\operatorname{GraphConv}(X_1) W_{\text {out }}+b_{out})+X,\end{equation}
where $Y$ has the same size as $X$, $W_{\text {in }}$ and $W_{\text {out }}$ are the weights. The activation function used is $\operatorname{GELU}$ \cite{hendrycks2016bridging}. $b_{in}$ and $b_{out}$ are biases. In the implementation, all $InputW+b$ are achieved by using a convolutional layer following a batch normalization operation. $\operatorname{GraphConv}$ means aggregating and updating the discussed graph-level processing. The Grapher module is with a shortcut structure. The $\operatorname{Droppath}$ \cite{larsson2016fractalnet} operation is used.

\subsection{Feed-forward Network}
Feed-forward Networks (FFNs) are used to help with the feature transformation capacity and relief the over-smoothing phenomenon after the Grapher module. The FFN can be represented as 
\begin{equation}Z=\operatorname{Droppath}(\operatorname{GELU}\left(Y W_1+b_1\right) W_2+b_2)+Y,\end{equation}
where $W_1$ and $W_2$ are weights, $b_1$ and $b_2$ are biases. In the implementation, each feed-forward operation $InputW+b$ is achieved by using a convolutional layer following a batch normalization operation. The $\operatorname{Droppath}$ operation is used. The workflow of Grapher and FFN modules is shown in Figure2.

\begin{figure}[htbp]
\centering
\centerline{\includegraphics[width=9cm]{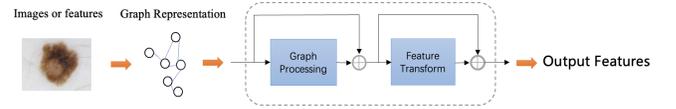}}
\caption{The workflow of Grapher and FFN modules: graph processing and feature transformation are applied}
\end{figure}

\begin{figure*}[htbp]
\centering
\centerline{\includegraphics[width=16.15cm]{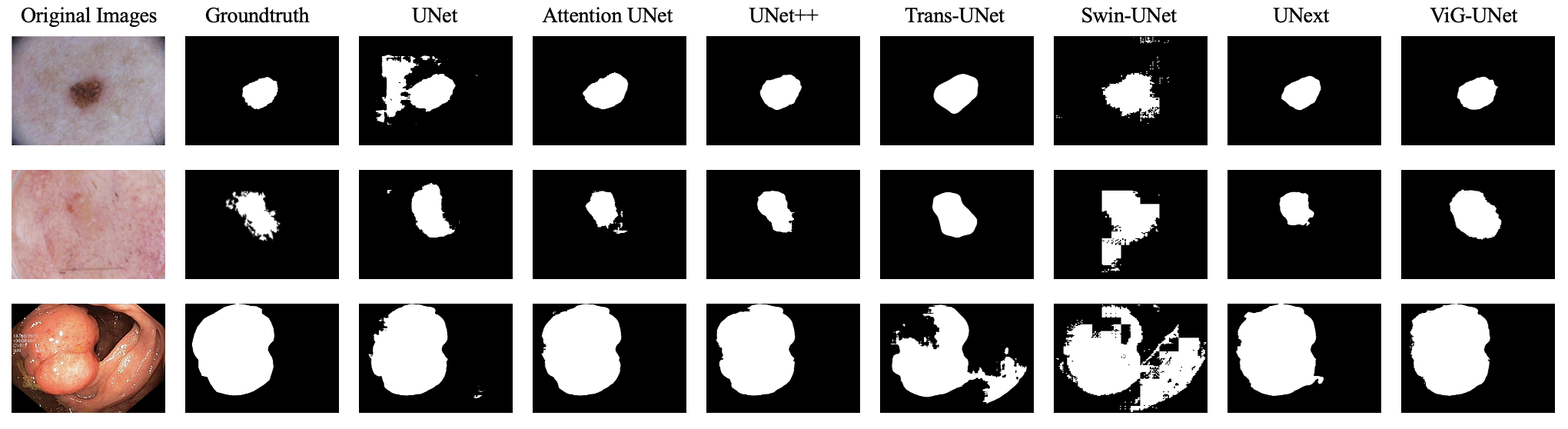}}
\caption{The example segmentation experimental results of different methods on ISIC 2016, ISIC 2017 and Kvasir-SEG datasets}
\end{figure*}
\section{Experiments}
\subsection{Datasets}
\textbf{ISIC 2016} is a dataset of dermoscopic images of skin lesions. We used the dataset of lesion segmentation in this paper. There are 900 pairs of images and corresponding masks in the training set. In the testing set, there are 379 pairs. \textbf{ISIC 2017} is a dataset of dermoscopic images of skin lesions. We used the dataset of the lesion segmentation task, which contains images and corresponding masks. There are 2000 pairs in the training set, 150 in the validation set, and 600 in the testing set. The \textbf{Kvasir-SEG} dataset contains 1000 pairs of polyp images and masks. We split the dataset into training and testing sets with a ratio of 0.2 with a random state of 41. 

\subsection{Implementation Details}
The experiments are all done on PG500-216(V-100) with 32 GB memory. The training and validation set of ISIC 2016 and Kvair-SEG are split with a ratio of 0.2 with the random state 41. The total training epochs are 200 and the batch size is 4. The input images are all resized to 512$\times$512. The optimizer used is ADAM \cite{kingma2014adam}. The initial learning rate is 0.0001 and a CosineAnnealingLR \cite{loshchilov2016sgdr} scheduler is used. The minimum learning rate is 0.00001. Only rotation by 90 degrees clockwise for random times, flipping and normalization methods are used for augmentation. The evaluation metrics in validation are $IOU$ of the lesions. A mixed loss combining binary cross entropy (BCE) loss and dice loss \cite{milletari2016v} is used in the training process:$$\mathcal{L}=0.5 B C E(\hat{y}, y)+D i c e(\hat{y}, y)$$ 

We implemented ViG-UNet and six other U-Net variants for comparison experiments. The pre-trained Swin-T model of $224\times 224$ input size on ImageNet 1k is used for Swin-UNet, and the pre-trained ViT-B/16 on ImageNet 21k is used for Trans-UNet.

\begin{table}[h]
\caption{Comparison Experimental Results on ISIC 2016, ISIC 2017 and Kvasir SEG (using the $IoU$ metric)}
\label{comp}
\center
 \setlength{\tabcolsep}{1mm}{
\begin{tabular}{cccc}
\hline
Methods                        &  ISIC 2016 &  ISIC 2017     & Kvasir SEG       \\ \hline
UNet \cite{ronneberger2015u}  & 0.8209  &  0.6410 & 0.6913    \\
Attention-UNet \cite{oktay2018attention} &  0.8325 & 0.6473  &  0.6946 \\
UNet++ \cite{zhou2018unet++}  & 0.8343& 0.6504 & 0.6906\\
Trans-UNet \cite{chen2021transunet} & 0.8481 &0.7147& 0.4943\\
Swin-UNet \cite{cao2021swin}  & 0.7559 & 0.6676& 0.3405\\
UNext \cite{valanarasu2022unext} & 0.8397 & 0.7156& 0.6996\\ \hline
ViG-UNet&              \textbf{0.8558} &\textbf{0.7211}& \textbf{0.7104}\\ \hline
\end{tabular}}
\end{table}

\begin{table}[h]
\caption{Comparison Experimental Results on ISIC 2016, ISIC 2017 and Kvasir SEG (using the $Dice$ metric)}
\label{comp}
\center
 \setlength{\tabcolsep}{1mm}{
\begin{tabular}{cccc}
\hline
Methods                        &  ISIC 2016 &  ISIC 2017     & Kvasir SEG       \\ \hline
UNet   & 0.8984  &  0.7708 & 0.8023    \\
Attention-UNet  &  0.9058 &  0.7739  &   0.8065 \\
UNet++  & 0.9070& 0.7768 & 0.8033\\
Trans-UNet & 0.9158 &0.8244& 0.6439\\
Swin-UNet   & 0.8568 &  0.7914& 0.4974\\
UNext  & 0.9103 & 0.8241& 0.8122\\ \hline
ViG-UNet&              \textbf{0.9206} &\textbf{0.8292}& \textbf{0.8188}\\ \hline
\end{tabular}}
\end{table}

\begin{table}[h]
\caption{Comparison of Parameters of Different Models}
\label{comp}
\center
 \setlength{\tabcolsep}{2mm}{
\begin{tabular}{ccc}
\hline
Methods                        &  Parameters        \\ \hline
UNet & 7.8M     \\
Attention-UNet  &  8.7M  \\
UNet++ & 9.2M\\
Trans-UNet  & 92.3M \\
Swin-UNet   & 27.3M \\
UNext & 1.5M  \\ \hline
ViG-UNet&              0.7G  \\ \hline
\end{tabular}}
\end{table}

\subsection{Results}
The performances of different methods are shown in Table 2 and Table 3 with metrics of $IoU$ and $Dice$. The example segmentation results of different methods are displayed in Figure 3. From these experiments, we can see that on all three datasets, our approach performs best. And we can expect that if we use pre-trained models of ViG on ImageNet, the performance may be better. For the Swin-UNet, it's strange, but \cite{valanarasu2022unext} also reports its low performance on small datasets. In conclusion, our method shows competitive performance compared to classical and state-of-the-art techniques. 

We also calculate the parameters by using fvcore Python package with (1, 3, 512, 512) input. Admittedly, our model is larger than others and needs more computational resources.

\section{Conclusion}
In this work, we propose a ViG-UNet for 2D medical image segmentation, which has a GNN-based U-shaped architecture consisting of the encoder, the bottleneck, and the decoder with skip connections. Experiments are done on the ISIC 2016, the ISIC 2017 and the Kvasir-SEG dataset, whose results show that our method is effective.


\section{Acknowledgements}
This work was supported by a Grant from The National Natural Science Foundation of China (No. U21A20484).
\bibliographystyle{IEEEbib}
\bibliography{refs}

\begin{thebibliography}{10}

\bibitem{ronneberger2015u}
Olaf Ronneberger et~al.,
\newblock ``U-net: Convolutional networks for biomedical image segmentation,''
\newblock in {\em International Conference on Medical image computing and
  computer-assisted intervention}. Springer, 2015, pp. 234--241.

\bibitem{oktay2018attention}
Ozan Oktay et~al.,
\newblock ``Attention u-net: Learning where to look for the pancreas,''
\newblock {\em arXiv preprint arXiv:1804.03999}, 2018.

\bibitem{zhou2018unet++}
Zongwei Zhou et~al.,
\newblock ``Unet++: A nested u-net architecture for medical image
  segmentation,''
\newblock in {\em Deep learning in medical image analysis and multimodal
  learning for clinical decision support}, pp. 3--11. Springer, 2018.

\bibitem{dosovitskiy2020image}
Alexey Dosovitskiy et~al.,
\newblock ``An image is worth 16x16 words: Transformers for image recognition
  at scale,''
\newblock {\em arXiv preprint arXiv:2010.11929}, 2020.

\bibitem{chen2021transunet}
Jieneng Chen et~al.,
\newblock ``Transunet: Transformers make strong encoders for medical image
  segmentation,''
\newblock {\em arXiv preprint arXiv:2102.04306}, 2021.

\bibitem{cao2021swin}
Hu~Cao et~al.,
\newblock ``Swin-unet: Unet-like pure transformer for medical image
  segmentation,''
\newblock {\em arXiv preprint arXiv:2105.05537}, 2021.

\bibitem{gori2005new}
Marco Gori et~al.,
\newblock ``A new model for learning in graph domains,''
\newblock in {\em Proceedings. 2005 IEEE international joint conference on
  neural networks}, 2005, vol.~2, pp. 729--734.

\bibitem{micheli2009neural}
Alessio Micheli,
\newblock ``Neural network for graphs: A contextual constructive approach,''
\newblock {\em IEEE Transactions on Neural Networks}, vol. 20, no. 3, pp.
  498--511, 2009.

\bibitem{niepert2016learning}
Mathias Niepert et~al.,
\newblock ``Learning convolutional neural networks for graphs,''
\newblock in {\em International conference on machine learning}. PMLR, 2016,
  pp. 2014--2023.

\bibitem{bruna2013spectral}
Joan Bruna et~al.,
\newblock ``Spectral networks and locally connected networks on graphs,''
\newblock {\em arXiv preprint arXiv:1312.6203}, 2013.

\bibitem{henaff2015deep}
Mikael Henaff et~al.,
\newblock ``Deep convolutional networks on graph-structured data,''
\newblock {\em arXiv preprint arXiv:1506.05163}, 2015.

\bibitem{defferrard2016convolutional}
Micha{\"e}l Defferrard, Xavier Bresson, and Pierre Vandergheynst,
\newblock ``Convolutional neural networks on graphs with fast localized
  spectral filtering,''
\newblock {\em Advances in neural information processing systems}, vol. 29,
  2016.

\bibitem{kipf2016semi}
Thomas~N Kipf and Max Welling,
\newblock ``Semi-supervised classification with graph convolutional networks,''
\newblock {\em arXiv preprint arXiv:1609.02907}, 2016.

\bibitem{han2022vision}
Kai Han et~al.,
\newblock ``Vision gnn: An image is worth graph of nodes,''
\newblock {\em arXiv preprint arXiv:2206.00272}, 2022.

\bibitem{krizhevsky2017imagenet}
Alex Krizhevsky, Ilya Sutskever, and Geoffrey~E Hinton,
\newblock ``Imagenet classification with deep convolutional neural networks,''
\newblock {\em Communications of the ACM}, vol. 60, no. 6, pp. 84--90, 2017.

\bibitem{gutman2016skin}
David Gutman et~al.,
\newblock ``Skin lesion analysis toward melanoma detection: A challenge at the
  international symposium on biomedical imaging (isbi) 2016, hosted by the
  international skin imaging collaboration (isic),''
\newblock {\em arXiv preprint arXiv:1605.01397}, 2016.

\bibitem{codella2018skin}
Noel~CF Codella et~al.,
\newblock ``Skin lesion analysis toward melanoma detection: A challenge at the
  2017 international symposium on biomedical imaging (isbi), hosted by the
  international skin imaging collaboration (isic),''
\newblock in {\em 2018 IEEE 15th international symposium on biomedical imaging
  (ISBI 2018)}. IEEE, 2018, pp. 168--172.

\bibitem{jha2020kvasir}
Debesh Jha et~al.,
\newblock ``Kvasir-seg: A segmented polyp dataset,''
\newblock in {\em International Conference on Multimedia Modeling}. Springer,
  2020, pp. 451--462.

\bibitem{li2019deepgcns}
Guohao Li et~al.,
\newblock ``Deepgcns: Can gcns go as deep as cnns?,''
\newblock in {\em Proceedings of the IEEE/CVF international conference on
  computer vision}, 2019, pp. 9267--9276.

\bibitem{hendrycks2016bridging}
Dan Hendrycks and Kevin Gimpel,
\newblock ``Bridging nonlinearities and stochastic regularizers with gaussian
  error linear units,''
\newblock 2016.

\bibitem{larsson2016fractalnet}
Gustav Larsson, Michael Maire, and Gregory Shakhnarovich,
\newblock ``Fractalnet: Ultra-deep neural networks without residuals,''
\newblock {\em arXiv preprint arXiv:1605.07648}, 2016.

\bibitem{loshchilov2016sgdr}
Ilya Loshchilov and Frank Hutter,
\newblock ``Sgdr: Stochastic gradient descent with warm restarts,''
\newblock {\em arXiv preprint arXiv:1608.03983}, 2016.

\bibitem{milletari2016v}
Fausto Milletari et~al.,
\newblock ``V-net: Fully convolutional neural networks for volumetric medical
  image segmentation,''
\newblock in {\em 2016 fourth international conference on 3D vision (3DV)}.
  IEEE, 2016, pp. 565--571.

\bibitem{valanarasu2022unext}
Jeya Maria~Jose Valanarasu and Vishal~M Patel,
\newblock ``Unext: Mlp-based rapid medical image segmentation network,''
\newblock {\em arXiv preprint arXiv:2203.04967}, 2022.

\end{thebibliography}

\end{document}